# Giant Single-Molecule Magnets: A Mn$_{84}$ Torus and its Supramolecular Nanotubes[**]

Anastasios J. Tasiopoulos, Alina Vinslava, Wolfgang Wernsdorfer, Khalil A. Abboud and George Christou*

The discovery that individual molecules can function as magnets provided a new, 'bottom-up' approach to nanoscale magnetic materials,[1-3] and such molecules have since been called single-molecule magnets (SMMs).[4] Each molecule is a single-domain magnetic particle that, below its blocking temperature, exhibits the classical macroscale property of a magnet, namely magnetization hysteresis. In addition, SMMs straddle the classical/quantum interface in also displaying quantum tunnelling of magnetization (QTM)[5,6] and quantum phase interference,[7] the properties of the microscale. SMMs have various potential applications, including very high density information storage with each bit stored as the magnetization orientation of an individual molecule, and as quantum bits for quantum computing,[8] taking advantage of the quantum superposition of states provided by the QTM. For a number of reasons, including facilitating development of techniques for addressing individual SMMs, we have sought to synthesize SMMs of very large dimensions (by molecular standards). In effect, can the molecular (or 'bottom-up') approach reach the size regime of the classical (or 'top-down') approach to nanoscale magnetic materials? Indeed, here we report a giant $Mn_{84}$ SMM. It possesses a 4 nm diameter torus structure, exhibits both magnetization hysteresis and QTM, and crystallizes as supramolecular nanotubes.

[*] Prof. Dr. G. Christou, Dr. Anastasios J. Tasiopoulos, Dr. Khalil A. Abboud, Alina Vinslava.
Department of Chemistry
University of Florida
Gainesville, Florida 32611-7200, USA
Fax: (+1) 352-392-8757
E-mail: christou@chem.ufl.edu
Dr. Wolfgang Wernsdorfer
Laboratoire Louis Néel-CNRS
BP 166, 25 Avenue des Martyrs
38042 Grenoble, Cedex 9, France

[**] This work was supported by the U.S. National Science Foundation (Grant CHE-0123603). We thank Philip C. E. Stamp for useful discussions.

The compound $[Mn_{84}O_{72}(O_2CMe)_{78}(OMe)_{24}(MeOH)_{12}(H_2O)_{42}(OH)_6]·xH_2O·yCHCl_3$ (**1**·$xH_2O·yCHCl_3$) was obtained from the reaction of $[Mn_{12}O_{12}(O_2CMe)_{16}(H_2O)_4]·4H_2O·2MeCO_2H$ (**2**) with $(NBu^n_4)(MnO_4)$ in MeOH, containing a little acetic acid, followed by filtration and layering of the filtrate with chloroform. After several weeks, the well-formed reddish-brown crystals were isolated in a 20% overall yield, based on total available Mn. The compound crystallizes in hexagonal space group *P*6 with the asymmetric unit containing 1/6 of the molecule and approximately 25 water and 2 chloroform solvent molecules of crystallization, which are severely disordered. The structure[9] (Figure 1) comprises a $Mn_{84}$ torus with $C_6$ crystallographic symmetry composed of alternating near-linear $[Mn_3O_4]$ and cubic $[Mn_4O_2(OMe)_2]$ sub-units. All the metal centres are six-coordinate. Close inspection of Mn-O bond lengths, Mn bond valence sum calculations (2.82 – 3.26),[10] and the detection of Jahn-Teller axial elongations, as expected for a $d^4$ metal ion in near-octahedral geometry, identified the metal ions as all being in the $Mn^{3+}$ oxidation state.

A better appreciation of the structure and the size of the molecule is provided by the space-filling plots of Figure 2, which show that the torus has a diameter of ~ 4.2 nm and a thickness of ~1.2 nm, with a central hole of diameter 1.9 nm. The $Mn_{84}$ molecules order within the crystal in an aesthetically very pleasing manner, giving nanotubular stacks parallel to the crystal c axis and to their neighbours (Figs 2b,c). This yields a hexagonal close packing analogous to densely packed straws in a box. The crystalline structure of $Mn_{84}$ thus displays extensive cylindrical channel formation along one dimension. Molecules in neighbouring chains are exactly adjacent (Fig 2b), and thus the structure may be alternatively described as consisting of graphite-like $Mn_{84}$ sheets lying on top of each other with perfect registry.

The magnetic properties of $Mn_{84}$ have been investigated by both DC and AC methods. The molecule comprises eighty-four $Mn^{3+}$ (S = 2) ions, and it was anticipated that there would be fairly strong pairwise exchange interactions between them because they are all mono-atomically bridged by $O^{2-}$ or $MeO^-$ groups. Preliminary magnetic susceptibility ($\chi_M$) studies on polycrystalline samples down to 1.8 K indicated a small uncompensated molecular spin of $S \approx 6$. In the presence of sufficient magnetic anisotropy of the easy axis (Ising) type, a spin of this magnitude is sufficient to provide a SMM, and more detailed studies to lower temperatures were therefore carried out to investigate this possibility.

Studies were performed by magnetization measurements on single crystals of $Mn_{84}$ using an array of micro-SQUIDs.[11] The magnetization vs applied DC field at a 0.035 T/s sweep rate (Fig. 3a) exhibited hysteresis, the diagnostic property of a magnet, establishing $Mn_{84}$ as the largest SMM yet discovered. At this sweep rate, the hysteresis becomes evident at 1.5 K, and its coercivity increases with decreasing temperature, as expected for the superparamagnet-like properties of a SMM, before becoming essentially temperature-independent below 0.3 K. The magnetization in Fig. 3a is plotted as spin ($S$) per molecule (determined using quantitative molar magnetization data). The saturation value indicates a molecular ground state spin of $S = 6$, confirming the preliminary estimate above. The hysteresis loop shows no sign of the steps diagnostic of QTM that are visible in the hysteresis loops of several smaller SMMs such as $Mn_{12}$,[5, 6] $Mn_4$,[12, 13] and $Fe_8$,[7, 14] and the exchange-biased $[Mn_4]_2$ dimer.[15] In previous cases where no steps are visible, for large SMMs such as the $[Mn_{18}]^{2+}$ [16] and $Mn_{30}$ [17] SMMs, it is due to a broadening and smearing of the steps from a distribution of molecular environments (and thus a distribution of relaxation barriers) caused by disordered lattice solvent molecules and ligand disorder; the magnetic properties of SMMs are sensitive to such relatively small variations in local environments. For $Mn_{84}$, the large numbers of disordered solvent molecules in the central cavity readily rationalize a distribution of molecular environments.

With Fig. 3a establishing $Mn_{84}$ as a SMM but not showing steps, it became of importance to establish via an alternative means whether quantum tunnelling is still occurring in such a giant SMM. This was addressed by collecting magnetization relaxation data from (i) out-of-phase AC susceptibility ($\chi_M''$) data at different AC frequencies, and (ii) magnetization vs time decay plots. Both methods provide magnetization relaxation rate ($1/\tau$) vs. $T$ data, where $\tau$ is the relaxation time. These data are plotted as $\tau$ vs $1/T$ in Fig. 3b. There is a thermally activated region above ~ 0.5 K, and fitting of this region to the Arrhenius relationship of Equation (1) gives $U_{eff}$ = 18 K and $\tau_0 = 5.7 \times 10^{-9}$ s, where $U_{eff}$ is the effective (mean) relaxation

$$(1/\tau) = (1/\tau_0) \exp(-U_{eff}/kT) \qquad (1)$$

barrier and $\tau_0$ is the pre-exponential factor. The relaxation rate levels off below 0.5 K, and below ~0.2 K it is temperature-independent, consistent with relaxation only by ground state tunnelling between the lowest energy $M_S = \pm 6$ levels of the $S = 6$ manifold. Such temperature-independent relaxation rates due to QTM have been observed previously for several other SMMs.[12-14, 16-18] Independent conformation of QTM in $Mn_{84}$ was obtained from 'quantum hole digging',[19] which can establish QTM even when no steps are apparent in hysteresis loops. It is thus clear that $Mn_{84}$ still exhibits the quantum behaviour that has become a common feature of the much smaller SMMs.

The above results establish $Mn_{84}$ as a giant SMM. In general, SMMs possess many advantages over classical magnetic particles made of Co metal, $Fe_3O_4$, etc., including monodispersity, crystallization as highly ordered ensembles, true solubility (rather than colloidal suspensions), and a shell of organic groups that protects the magnetic cores from surface variations and prevents their contact. The discovery of a $Mn_{84}$ SMM establishes that these advantages can also be extended to this giant molecule, which is still soluble, stable, and crystalline. Such large wheel-like molecules are with precedent, but only in molybdenum chemistry,[20] with the largest currently known being the $Mo_{154}$[21] and $Mo_{176}$[22] compounds prepared by Müller and coworkers. Although they contain more metal atoms than $Mn_{84}$, the $Mo_{154}$ and $Mo_{176}$ wheels have diameters of ~3.4 and ~ 4.1 nm respectively. These and other giant polyoxometallates,[23-25] are diamagnetic, or nearly so, and none of them has thus been found to exhibit SMM properties.

But although $Mn_{84}$ is very large by molecular standards, it is pertinent to ask how it compares to classical nanoparticles. In Fig. 4 are compared the sizes of $Mn_4$, $Mn_{12}$, $Mn_{30}$ and $Mn_{84}$ SMMs with that of a 3 nm Co nanoparticle recently reported in the literature,[26] all drawn to the same scale. With a 4.2 nm diameter, $Mn_{84}$ is thus of comparable size to the smallest nanoparticles. The total number of atoms in $Mn_{84}$ is 1032, the same as the 3 nm Co nanoparticle, which contains ~1000 Co atoms. Of course, the former has a very different shape, given its central hole and essentially wheel-like rather than spherical structure. A useful additional way, particularly in the physics literature, of comparing the 'size' of magnetic systems is by their Néel vector ($N$, the sum of the individual spins), which are 7.5, 22, 61 and 168 for $Mn_4$, $Mn_{12}$, $Mn_{30}$ and $Mn_{84}$, respectively. This is the scale used in Fig. 4. With a value of 168, $Mn_{84}$ is a far larger spin system than any other molecular cluster (greater than the next largest 75 observed previously for an $Fe_{30}$ species with S = 0[24, 25]) and is at the lower limit of values found for classical nanoparticles, which can range from a few hundred to many thousands depending on the precise size and constituent metal; for the 3 nm Co nanoparticle shown, the Néel vector is approximately 1000.

The new $Mn_{84}$ SMM thus essentially represents a long sought-after meeting of the bottom-up and top-down approaches to nanoscale magnetic materials. It will be interesting to see whether even larger SMMs containing more metal atoms can be obtained in the near future; we see no reason why this should not be so. It is also interesting that $Mn_{84}$ exhibits magnetization quantum tunnelling, a point of relevance to the general question of how large magnetic nanoparticles can become and still display quantum effects.[27] Such studies for classical magnetic nanoparticles have been hampered by complications from distributions of particle size, shape, surface roughness, and spin[11, 28] a problem not existing for crystalline, monodisperse SMMs. Finally, we recognize that the supramolecular architecture of the $Mn_{84}$ molecules as ordered nanotubular stacks offers a variety of future possibilities for accessing interesting new materials, including the insertion of guest molecules (either organic or inorganic, diamagnetic or paramagnetic) or chains (either conducting or insulating) into or through the tubes. In addition, the large shape anisotropy and solubility of $Mn_{84}$ suggest that it should be fairly straightforward to deposit and visualize these molecular rings on surfaces.

*Experimental Section*

Compound **1**: To a slurry of [Mn$_{12}$O$_{12}$(O$_2$CMe)$_{16}$(H$_2$O)$_4$]·4H$_2$O·2HO$_2$CMe (0.425 g, 0.206 mmol) in MeOH (15 ml) was added over a period of two minutes a freshly-prepared solution of NBu$^n_4$MnO$_4$ (0.30 g, 0.83 mmol) in MeOH (10 ml) and glacial acetic acid (0.75ml). The mixture was left under magnetic stirring for a few minutes, and then filtered to give a reddish-brown filtrate. Equal amounts of the filtrate were placed in five different vials and layered with chloroform. After a few weeks reddish-brown crystals appeared, and were left to grow for several more weeks to give well shaped X-rays quality crystals. When product formation was deemed to be complete, the crystals were collected by filtration, washed with chloroform and dried in vacuum. The yield was 0.10 g (20% based on total Mn). Elemental analysis: (%) calcd for C$_{192}$H$_{464}$O$_{322}$Mn$_{84}$ (**1**·10H$_2$O): C 18.39, H 3.73; found C 18.35, H 3.45. Selected IR data (KBr pellet): $\tilde{v}$ =3400(s, br), 2965(w), 2931(w), 1576(s, br), 1527(s, br), 1425(s, br), 1348(m), 1250(w), 1028(m), 949(w), 696(m), 666(s), 623(s), 580(s), 548(s), 506(m).

**Figure Captions.**

**Figure 1.** The structure of the $Mn_{84}$ torus, excluding hydrogen atoms. The rectangle shows the repeating $Mn_{14}$ unit that represents the contents of the asymmetric unit; for clarity, it is reproduced above the structure without the carbon atoms (except for those of the MeO⁻ groups to clarify their positions in the $Mn_{14}$ unit). Colour code: blue Mn; red O; grey C.

**Figure 2.** Space-filling representations of $Mn_{84}$, and its supramolecular aggregation into ordered nanotubes and sheets. (a) Space-filling representations (including hydrogen atoms) from viewpoints perpendicular (top) and parallel (bottom) to the plane of the torus, showing the dimensions of the molecule and its central hole. (b) Ordered arrangement of $Mn_{84}$ molecules (excluding hydrogen atoms) into two adjacent supramolecular nanotubes viewed perpendicular to the propagation axis; this view emphasizes the exact registry of molecules in adjacent tubes and thus the sheet-like structure formed. (c) A view along the propagation axes of seven tubes showing the hexagonal packing of neighbours within a single sheet of molecules (excluding hydrogen atoms). Colour code: blue Mn; red O; grey C; white H.

**Figure 3.** The results of magnetic susceptibility studies on single crystals of $Mn_{84}$. (a) Magnetization vs applied DC magnetic field plots, with the field applied along the c axis (perpendicular to the $Mn_{84}$ torus plane), exhibiting hysteresis loops. The magnetization has been plotted as spin ($S$) per $Mn_{84}$ molecule. A background slope due to low-lying excited states has been subtracted from the data. (b) Arrhenius plot constructed using a combination of out-of-phase AC susceptibility ($\chi_M''$) data and DC magnetization decay data. The dashed line is a fit of the thermally activated region to the Arrhenius relationship; see the text for the fit parameters.

**Figure 4**. The position of Mn$_{84}$ on a size scale spanning atomic to nanoscale dimensions. On the far right is shown a high-resolution transmission electron microscopy view along a [110] direction of a typical 3 nm diameter cobalt nanoparticle exhibiting a f.c.c. structure and containing ~1000 Co atoms.[26] The Mn$_{84}$ molecule is a 4.2 nm diameter particle. Also shown for comparison are the indicated smaller Mn SMMs, which are drawn to scale. An alternative means of comparison is the Néel vector ($N$), which is the scale shown. The green arrows indicate the magnitude of the Néel vectors for the indicated SMMs, which are 7.5, 22, 61 and 168 for Mn$_4$, Mn$_{12}$, Mn$_{30}$ and Mn$_{84}$, respectively. The green arrows from the Co nanoparticle are merely meant to indicate that the Néel vector of nanoparticles can take many values, depending on the exact size and the identity of the constituent metal.

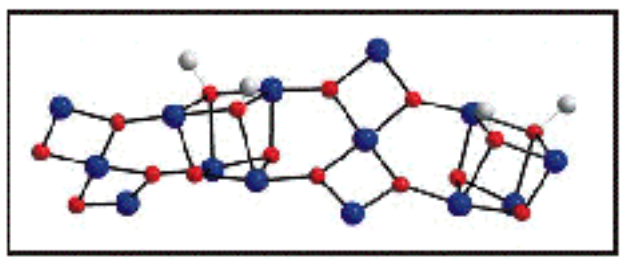
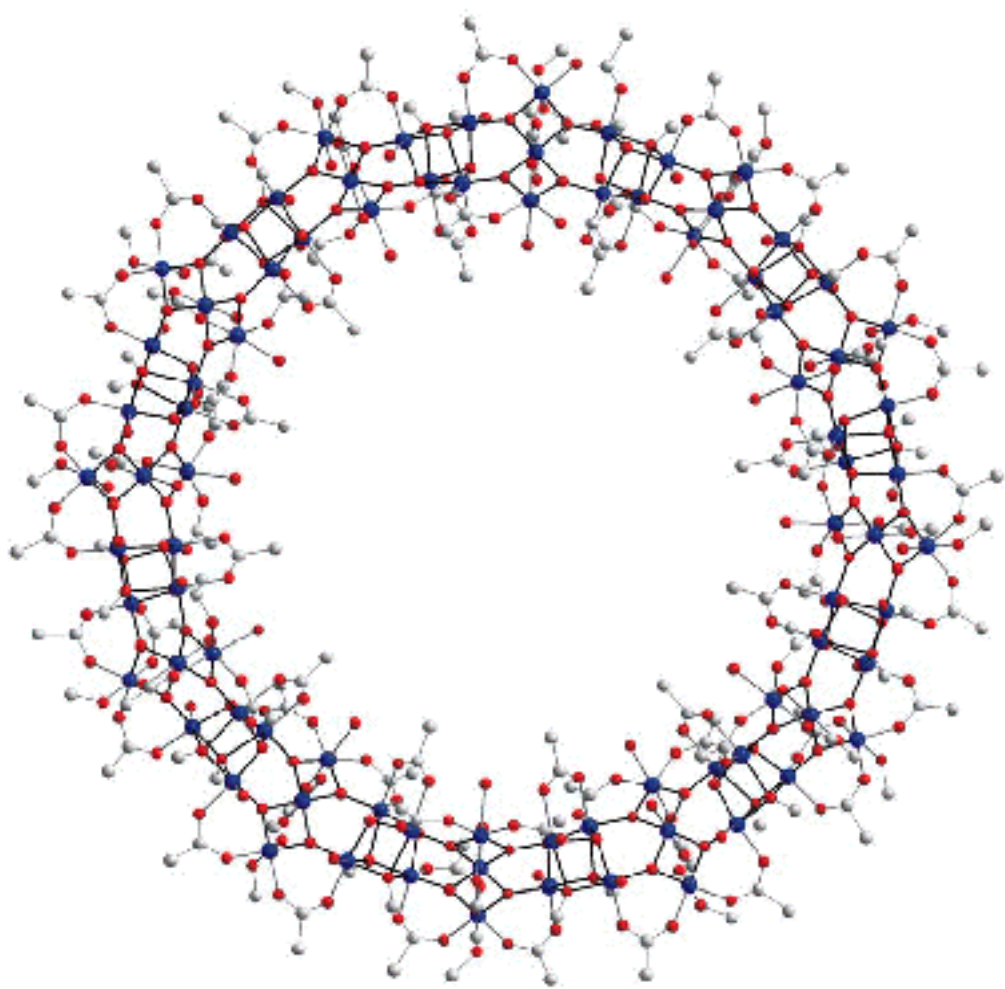

**Fig. 1**

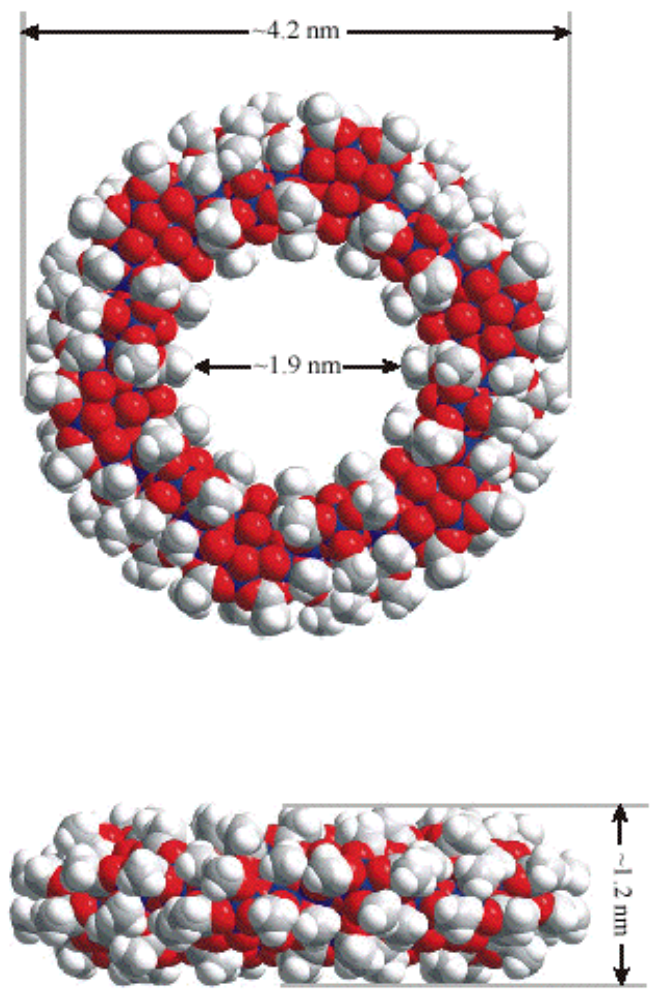

Fig. 2a

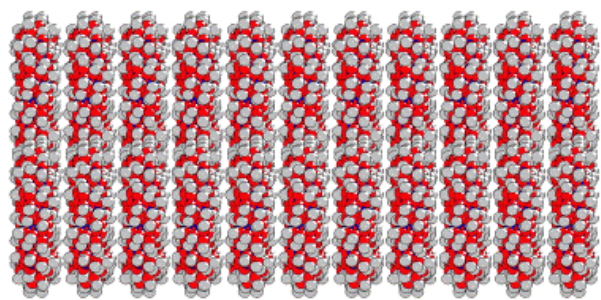

Fig. 2b

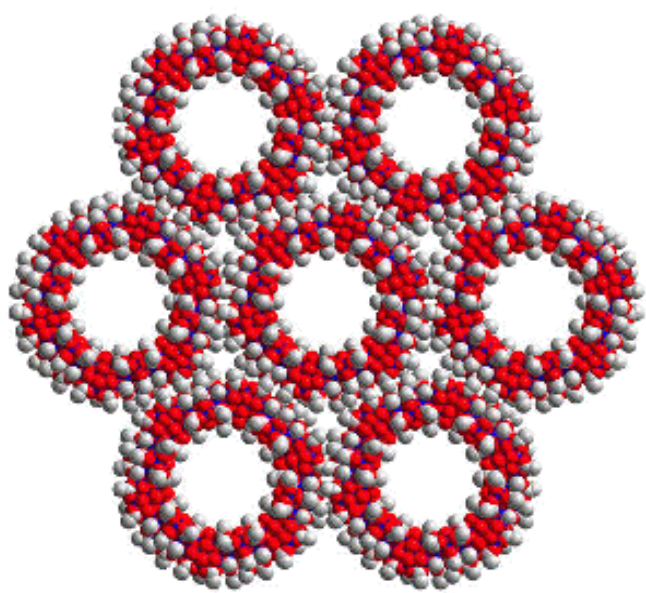

Fig. 2c

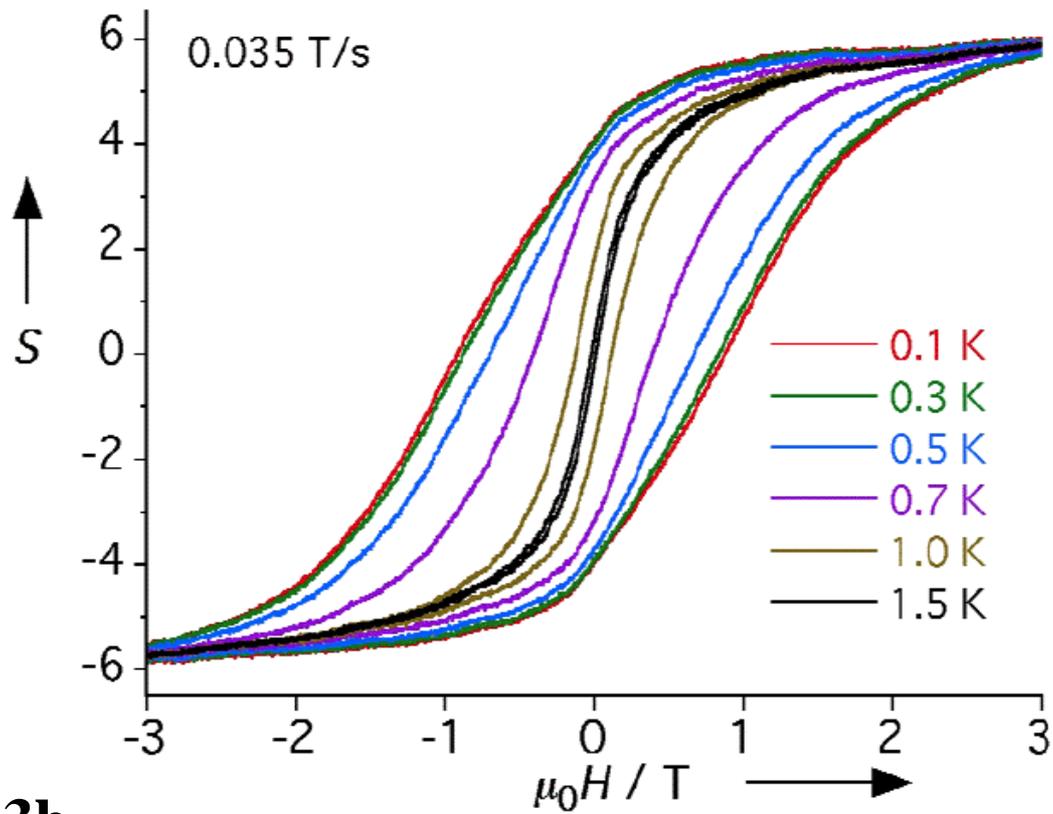

**Fig. 3a**

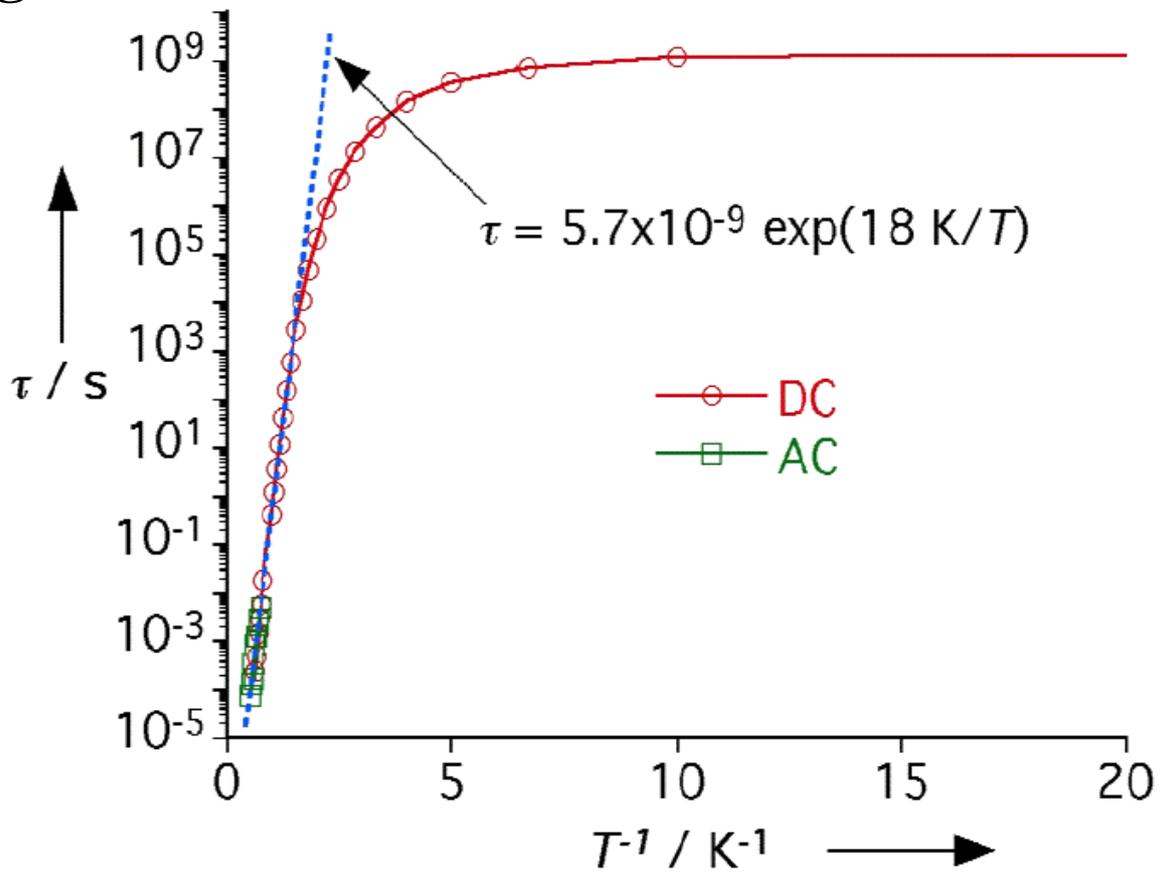

**Fig. 3b**

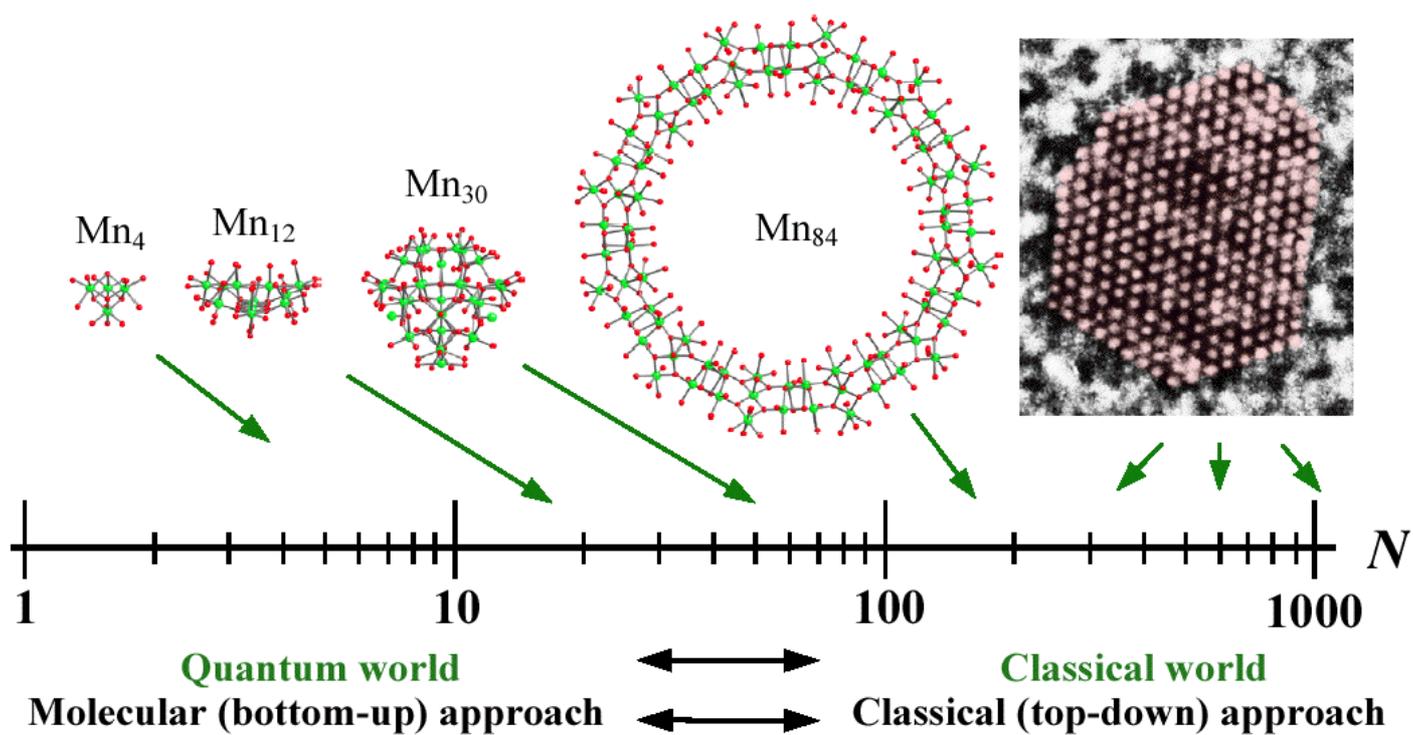

**Fig. 4**